# Deeply nonlinear excitation of self-normalised exchange spin waves

Qi Wang[1,2,3], Roman Verba[4], Björn Heinz[5], Michael Schneider[5], Ondřej Wojewoda[6], Kristýna Davídková[6], Khrystyna Levchenko[1], Carsten Dubs[7], Norbert J. Mauser[2,3], Michal Urbánek[6], Philipp Pirro[5], Andrii V. Chumak[1,2]

[1] *Faculty of Physics, University of Vienna, Vienna, Austria*

[2] *Research Platform Mathematics-Magnetism-Materials, Faculty of Math, University of Vienna, Vienna, Austria*

[3] *Wolfgang Pauli Institute, Vienna, Austria*

[4] *Institute of Magnetism, Kyiv, Ukraine*

[5] *Fachbereich Physik and Landesforschungszentrum OPTIMAS, Technische Universität Kaiserslautern, Kaiserslautern, Germany*

[6] *CEITEC BUT, Brno University of Technology, Brno, Czech Republic*

[7] *INNOVENT e.V., Technologieentwicklung, Jena, Germany*

**Abstract:**

Spin waves are ideal candidates for wave-based computing, but the construction of magnetic circuits is blocked by a lack of an efficient mechanism to excite long-running exchange spin waves with normalised amplitudes. Here, we solve the challenge by exploiting the deeply nonlinear phenomena of forward-volume spin waves in 200 nm wide nanoscale waveguides and validate our concept with microfocused Brillouin light scattering spectroscopy. An unprecedented nonlinear frequency shift of >2 GHz is achieved, corresponding to a magnetisation precession angle of 55° and enabling the excitation of exchange spin waves with a wavelength of down to ten nanometres with an efficiency of >80%. The amplitude of the excited spin waves is constant and independent of the input microwave power due to the self-locking nonlinear shift, enabling robust adjustment of the spin wave amplitudes in future on-chip magnonic integrated circuits.

**Introduction**

Magnonics is an emerging field of solid-state physics in which magnons, the quanta of spin waves, are used in place of electrons for information transmission and processing [1-5]. Spin waves, the collective excitations of the magnetic orders, provide a scalable wavelength [6-11] and exhibit a variety of distinct nonlinear phenomena [12-16] that make spin waves promising for Boolean computing [17-21], radio frequency applications [22,23] and neuromorphic computing [24-26] at the nanoscale. Beside their intrinsic nonlinearity, another advantage of spin waves are their short wavelengths down to a few nanometres [6-11], which makes it possible to design magnonic devices comparable in size to modern electronic devices using CMOS (complementary metal–oxide–semiconductor) technologies. The most common method to excite spin waves is to use an oscillating Oersted field generated by an alternating current in an inductive antenna. For a linear excitation, the width of the antenna limits the minimum wavelength of the excited spin waves. Therefore, the most straightforward way to excite short spin waves is to use nanoscale antennas. However, the spin-wave excitation efficiency of nano-antennas is greatly reduced due to the scaling and the increase in Ohmic resistance. Recently, several other methods have been developed for the excitation of short-wavelength exchange spin waves, e.g., using magnonic grating couplers [6], magnetic vortex cores [7, 27], parametric pumping [28, 29], and geometry-induced wavenumber convertors [30,31]. Most of these methods, however, have drawbacks such as selective excitation wavelengths, complex spin-wave emissions, relatively low excitation efficiency, or unrealistic integration. More importantly, the excited spin-wave power depends on the input microwave power in all the previously mentioned methods. This power-dependent excitation becomes a major obstacle for applications in large magnonic circuits, in which potentially millions of antennas need to be fabricated to feed the individual magnonic elements. Due to variations in the fabrications process, it is unrealistic and much too costly to precisely set the microwave power flowing in each antenna. Yet amplitudes that are not precisely set would significantly complicate the functioning of most linear and nonlinear concepts of wave logic such as the majority function or the half-adder. Therefore, developing a method to efficiently excite spin waves with amplitude self-normalisation is an important step toward the realisation of implementable magnonic circuits.

Here, we present a concise approach to excite spin waves with normalised amplitudes and with wavelengths down to tens of nanometres in nanoscale waveguides using a common microscopic inductive antenna with a width of 2 μm see Fig. 1. Such an antenna has low resistance, high spin-wave excitation efficiency and can be easily patterned using simple photolithography. The physical concept of the proposed method is based on the deeply nonlinear phenomena of forward volume spin waves (FVSWs) excited in normally-magnetised nanoscale waveguides [32-34] and allowed for controllable long-distance magnon transport.

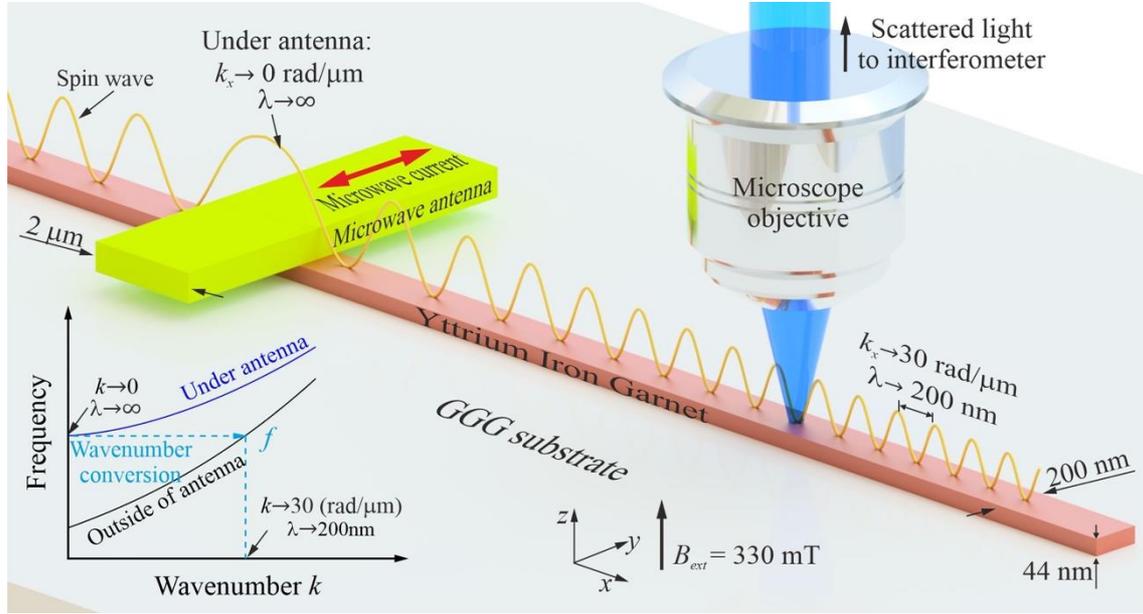

Figure 1. **The schematic picture of exchange spin-wave excitation.** Sketch of the sample and the experimental configuration: an inductive antenna with 2 μm width is placed on a nanoscale YIG waveguide with 200 nm width to excite the FMR mode under the antenna. μBLS spectroscopy is used to measure the spin waves along the $x$-axis of the nanoscale waveguide. The concept of the proposed method is based on the deep nonlinear shift of spin-wave dispersion in a normally magnetised nanoscale waveguide. The inset shows the schematic dispersion curves underneath the antenna and outside the antenna and the wavelength conversion from a few micrometres to 200 nm with spin waves of the excitation frequency $f$.

**Results**

**General concept of nonlinear exchange spin-wave excitation.** Figure 1 shows the schematic diagram of the general concept of our exchange spin-wave excitation experiments. A 200 nm wide yttrium iron garnet (YIG) waveguide is fabricated from a 44 nm-thin film using a hard-mask ion beam milling procedure (see Methods) [35,36]. A 2 μm-wide and 10 nm/150 nm-thick Ti/Au strip antenna is placed on top of the YIG waveguide to excite spin waves by sending a microwave current with a frequency $f$ and a power $P$. Note that the efficiently excited wavenumber range by this antenna is below $k_{max}=2\pi/w=3.14$ rad/μm ($w$ is the width of the antenna) in the linear region. An external field of 330 mT is applied out-of-plane along the $z$-axis, and forward volume spin waves are investigated. Microfocused Brillouin light scattering spectroscopy (μBLS) is employed to measure the spin-wave intensity as a function of the excitation frequency for different powers at different positions along the waveguide [35]. In this case of forward volume spin waves, the contribution of the linear magneto-optical coupling (i.e., Voigt effect) to the light scattering process is negligible and the second order magneto-optical coupling, known as the Cotton-Mouton effect, must be considered [37]. Thus, the BLS intensity is no longer proportional to the dynamic component of the out-of-plane magnetisation $\Delta m_z^2$, but is proportional to the dynamic components of the in-plane magnetisation $\Delta m_x^4$ and $\Delta m_y^4$ [37]. The inset of Fig. 1 illustrates the concept of the exchange spin wave excitation method based on the deep nonlinear frequency shift of FVSWs. A microwave current with a frequency $f$ above the linear ferromagnetic resonance (FMR) frequency $f_{\text{FMR}}$ ($\theta \rightarrow 0$) (where $\theta$ is precession angle) is sent to the antenna. For sufficient excitation powers, a nonlinear upshift of the dispersion relation as shown in the inset of Fig. 1 (blue curve) occurs, for instance, due to a forced excitation below the antenna. The FMR frequency $f_{\text{FMR}}$ of the spin waves in the region below the antenna shifts upwards to match the excitation frequency $f$, i.e., $f = f'_{\text{FMR}}$ ($\theta \gg 0$), which means that the excitation gets resonant. Due to the large size of the antenna, the microwave energy is efficiently pumped into the magnonic domain, i.e., the nonlinear FMR mode [38]. Consequently, as the spin waves propagating outside the antenna region have reduced amplitude, their dispersion curve quickly shifts back to lower frequencies due to the decreased nonlinear frequency shift. In the course of this shift, the energy of the FMR mode ($k\rightarrow 0$ rad/μm) is efficiently converted into exchange spin waves ($k\rightarrow 30$ rad/μm) as a result of the single-mode dispersion curve (black curve in the inset of Fig. 1) in the nanoscale waveguide [34,35].

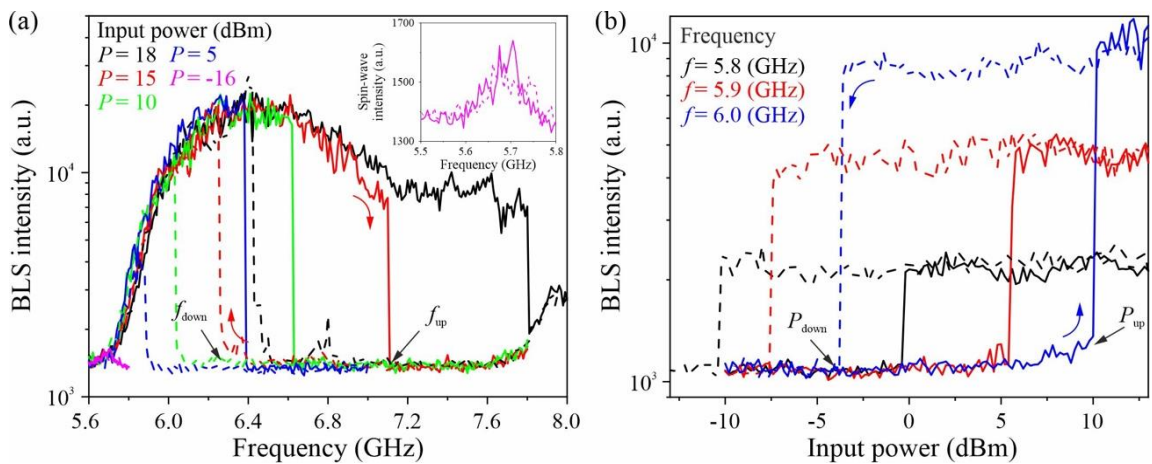

Figure 2. **Deep nonlinear shift, foldover effect and bistability.** (a) μBLS intensity as a function of the excitation frequency $f$ at different levels of the input power $P$. The inset shows the spin-wave spectra for input power of -16 dBm. The solid and dashed lines show the up- and down-sweep of the frequency,

respectively. (b) BLS intensity as a function of the input power *P* for different excitation frequencies *f*. The solid and dashed lines show the upward and downward power sweep.

**Deep nonlinear shift, foldover effect and bistability.** A continuously applied microwave current swept from 5.6 GHz to 8 GHz (or vice versa) with a step size of 10 MHz is sent to the strip antenna to excite spin waves. To detect all the excited spin waves (FMR mode and propagating spin waves), the focused laser beam of the μBLS is placed near the edge of the excitation antenna. Figure 2 shows the spin-wave spectra for different powers, and the solid and dashed lines show the frequency up- and down-sweep. For input powers lower than -16 dBm, the system operates in the linear region, and the upward and downward frequency sweep spectra almost overlap, as shown in the inset of Fig. 2(a). A small peak around 5.7 GHz is observed, indicating the ferromagnetic resonance (FMR) frequency. Micromagnetic simulations and analytic calculations show a similar FMR frequency of 5.76 GHz (see Supplementary Materials). For high input powers, the FMR frequency of the spin waves shifts upward with increasing excitation frequency due to the forced excitation causing a strong increase of the precession angle θ. As mentioned earlier, the BLS intensity of the FVSWs is proportional to $\Delta m_x^4$ and $\Delta m_y^4$ (i.e., $M_s \sin^4(\theta)$), and therefore the BLS intensity increases strongly with increasing precession angle (that in turn increases with the increase in frequency). The decrease in the BLS intensity in the high-frequency range (above 6.4 GHz) is attributable to the decrease of the detection efficiency of μBLS for short wavelengths [39,40]. Micromagnetic simulations excluding the influence of the detection efficiency show a monotonic increment in the spin-wave intensity with the increases in frequency (see Supplementary Materials). If the excitation frequency is increased further, at some point the microwave input power can no longer sustain the large precession of the magnetisation that would be necessary in order to shift the FMR frequency to the excitation frequency. At this point, the dispersion curve shifts back to the linear region. As a consequence, no spin wave can be excited and the BLS intensity drops to the noise level. Therefore, the spectrum exhibits a foldover behaviour: During the upward frequency sweep, the dispersion curve only needs to be shifted by a small frequency gap of 10 MHz at each step in our case, which merely requires an incremental increase of the already existing spin-wave intensity. However, in the downward sweep, the excitation frequency starts at a high frequency and moves to a low frequency. Since the system is initially not excited, the dispersion curve is in the linear region and the high excitation frequency is thus strongly off-resonant. As a result, the direct nonlinear excitation of the high-frequency spin waves is inefficient due to the large gap between the linear FMR frequency and the excitation frequency. When the excitation frequency decreases further, the microwave power can overcome the energy gap and the FMR frequency is nonlinearity shifted to the excitation frequency. Thus, a bistability window appears for the upward and downward frequency sweep as shown in Fig. 2(a). As expected from the discussion above, a higher RF power leads to a larger bistability window since larger precession angles and consequently a larger nonlinear shift can be sustained. Similar foldover and bistable spin-wave spectra are also observed in the micromagnetic simulations (see Supplementary Materials).

At an input power of 18 dBm, an unprecedented nonlinear frequency shift of 2.1 GHz (from 5.7 GHz to 7.8 GHz) and a large bistability window of 1.2 GHz (from 6.4 GHz to 7.8 GHz) are observed. This huge nonlinear frequency shift is two orders of magnitude larger than the frequency shift of tens of MHz noticed in previous studies [32,33,41]. Very recently, Merbouche *et al.* reported the self-phase modulated propagation of FVSWs in a 1 μm wide YIG waveguide with a precession angle around 12° and a 100 MHz nonlinear frequency shift [42]. We attribute the limitations of the observed nonlinear frequency shift in the previous studies to the onset of spin-wave instabilities caused by the interaction between different width/thickness modes in macro/micro-scale waveguides [32,33,41,42]. However, the width/thickness modes in our nanoscale waveguides are well separated due to the strong contribution of exchange energy that shifts the frequency of the higher-order width/thickness modes by several gigahertz. In our case, the third width mode (see Supplementary Materials) starts appearing in the spin-wave spectra at 7.7 GHz as illustrated in Fig. 2(a). Thus, the 200 nm-wide waveguide provides a frequency range of about 2 GHz for the single-mode dispersion curve, which is an intrinsic benefit of the nanoscale waveguides [34,35]. Note that the even width mode (antisymmetric mode) has a much lower excitation and conversion efficiency, compared to the odd modes, and can be ignored in our case. Further experiments using pulsed instead of continuous excitation indicate that the frequency shift caused by heating can be ignored and thus, the frequency shift is mainly caused by nonlinear phenomena (see Supplementary Materials).

Figure 2(a) also shows other important features: that the BLS intensity (proportional to the square of spin-wave intensity [37]) first increases with increasing frequency and, most importantly, does not depend on the input power. The independence of the spin-wave intensity on the input power over a wide range is validated by power-swept measurements for fixed frequencies. Figure 2(b) clearly shows that the excited spin-wave intensity is nearly constant (for a change of the input power by up to two orders of magnitude) once the input power is above a critical value. This behaviour that the output spin-wave power is independent of the input power can be used as a microwave power limiter and allows to create a self-normalising spin-wave source mandatory for many practical applications. From this observation, we can conclude that under the antenna only the FMR mode is excited. The precession amplitude, then, is determined mainly from the condition that the nonlinear FMR frequency matches the excitation frequency. In our geometry, the shifted frequency of the FMR mode can be described as

$$f(\bar{\theta}) = f_{\text{FMR}} + T_k(1 - \cos(\bar{\theta})) \qquad (1)$$

where $f_{\text{FMR}}$ is the FMR frequency in the linear region, $T_k$ is the nonlinear frequency shift coefficient, and $\bar{\theta}$ is the width-averaged precession angle (in our case, $f_{\text{FMR}}$=5.76 GHz, $T_k$=4.18 GHz, details see Methods). Then, $\cos(\bar{\theta}) = 1 - (f - f_{\text{FMR}})/T_k$, while the input power can cause only weak deviation (of the order of the FMR linewidth) from this condition to achieve the equilibrium between pumped and dissipated power. The critical power is the minimum power that can satisfy this balance (with zero detuning from the nonlinear resonance condition). Below the critical power, the drive is not strong enough to sustain a precession with the necessary angle and instead of a resonant, large-angle nonlinear excitation, only a weak, deeply non-resonant linear excitation takes place.

The self-normalised spin-wave emission is a unique feature of the nonlinear excitation and solves one of the critical challenges for the realisation of the magnonic circuits, in which potentially millions of antennas are to be fabricated in the circuits for different purposes. This method removes the need to precisely control microwave input power for each antenna (for instance, if the antennas have different resistivities, they still emit the same amount of spin waves on the applied frequency). The self-normalisation properties provided by the nonlinear phenomena will dramatically simplify the design of magnonic circuits.

**Spin-wave wavelength conversion.** Another advantage of the huge nonlinear frequency shift is that it provides a large wavenumber conversion up to 30 rad/μm in *k*-space and allows the excitation of spin waves with wavelengths of ~200 nm. Micromagnetic simulations are performed and a semi-analytical theory of linear and nonlinear FVSWs dispersion curves is presented for nanoscale out-of-plane magnetised waveguides to investigate the physics behind the conversion. To model our experiments, in the simulations we first calculate the dynamic field distribution of a 2-μm wide strip antenna with a current of 25 mA in the magneto-static approximation as shown in the top panel of Fig. 3(a), and then plug it into Mumax[3] with a varying microwave frequency *f* to excite spin waves. The details of the micromagnetic simulation are indicated in Methods. The middle panel of Fig. 3(a) shows the static (time-averaged) internal field distribution (black line) and precession angle (red line) for the excitation frequency of 7.55 GHz. The internal field underneath the antenna is much larger than the field outside of this region, which is due to the reduction of the demagnetisation field caused by the large precession angle. The increase in internal field, together with other less evident nonlinear mechanisms, shift the FMR frequency of the dispersion curve from 5.7 GHz (linear case) to 7.55 GHz (excitation frequency), as shown in Fig. 3(b) (blue line). This large frequency shift corresponds to a huge precession angle (~55°) calculated by Eq. (1), also validated by the micromagnetic simulations. The bottom panel of Fig. 3(a) depicts a snapshot of the simulated spin-wave amplitude $m_x$ (averaged along width direction) as a function of propagation distance in the YIG waveguide. It can be seen that a uniform precession FMR mode with an averaged precession angle of about 57° is excited under the antenna. The precession angle decreases to around 42° just outside the antenna (Fig. 3(a)), resulting in a downward shift of the dispersion curve and corresponding wavenumber conversion from the $k_1 \rightarrow 0$ (FMR mode, $\lambda_1 \rightarrow \infty$) to $k_2$=19.4 rad/μm ($\lambda_2$=323 nm) as shown in Fig. 3(b) (green line). This conversion is very efficient due to the single-mode waveguide, in which the conversion of spin waves to higher width/thickness modes is suppressed. The large drop of the precession angle just outside the antenna can be understood recalling that under the antenna two counter-propagating spin waves (despite of their almost vanishing wavenumber) coexist, while outside the antenna there is only one wave.

Furthermore, we estimate the energy transfer ratio $T_R$ from the FMR mode to exchange spin waves calculating the power of propagating spin waves and the power dissipated under the antenna region (see Methods and Supplementary Materials). It shows that the efficiency is more than 80% in our case and is thus already much higher than reported for other methods [6-11]. It should be noted that despite the similarity, such conversion is much less efficient if the spin-wave dispersion shift is governed not by nonlinearity, but by spatial modulation of the external field or magnetic parameters, as a spin-wave partially reflects from such a sharp boundary [11,30-31,43]. To check the contribution of the nonlinearity to the energy conversion process, we artificially introduce a spatially dependent internal field (similar to the middle panel of Fig. 3(a)) into Mumax$^3$ and excite a linear spin wave (precession angle <1°) at the same frequency of 7.55 GHz. Similar wavelength conversion is observed, but with lower energy transmission efficiency (~55%) (see Supplementary Materials). If we compare the parasitic losses ($L_p = 1 - T_R$), the nonlinear mechanism is around 2.6 times more efficient than that in the linear case. Therefore, the nonlinear self-sustained mechanism presented here makes a large contribution to the energy conversion from FMR mode to exchange spin waves. Another advantage of this mechanism is the short conversion distance, i.e., the short wavelength of spin waves directly occurs in the immediate vicinity of the antenna at a short conversion distance.

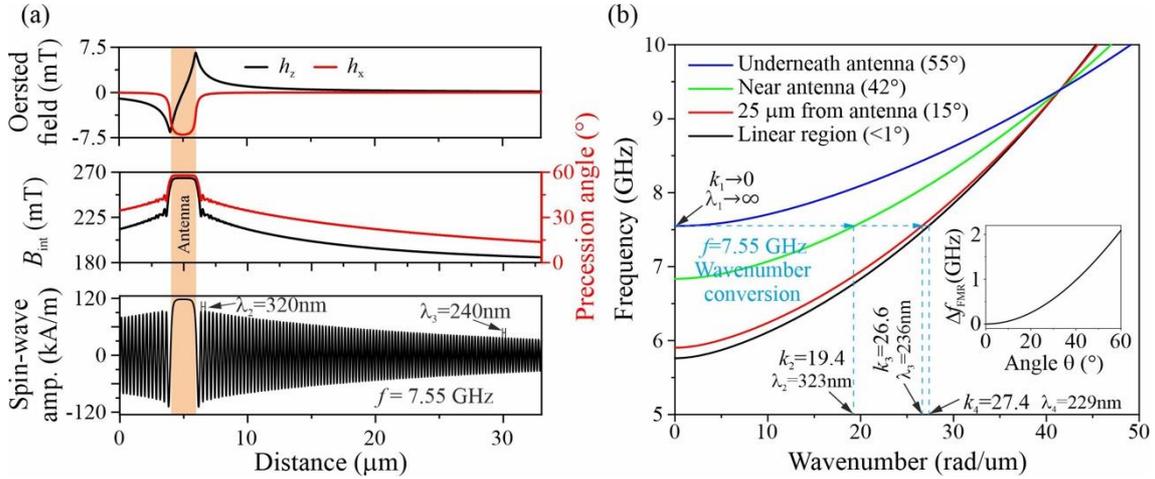

Figure 3. **Physical origin of spin-wave wavelength conversion.** (a) Top panel: the calculated Oersted field distribution caused by the microwave current in a 2 μm-wide strip antenna with current of 25 mA. Middle panel: the time-averaged internal field (black line) and precession angle (red line) for spin waves of frequency 7.55 GHz in the YIG waveguide extracted from Mumax[3]. Bottom panel: the simulated width-averaged spin-wave amplitude ($m_x$) as a function of the propagation distance in the YIG waveguide. (b) The analytical calculation of the dispersion curves in the YIG waveguide for different precession angles and the principle of the wavelength conversion.

Subsequently, the amplitude of the propagating spin wave decreases along the propagation direction due to the Gilbert damping, and thus, the internal field decreases smoothly. The bottom panel of Fig. 3(a) shows that the precession angle of spin waves reduces to around 15° after 25 μm propagation distance and the wavelength decreases further to $\lambda_3$=240 nm. This conversion in $k$ space is also shown in Fig. 3(b) between the green and red lines. Finally, the dispersion curve shifts back to the linear region (precession angle <1°) after several tens of micrometres propagation distance, as depicted in Fig. 3(b) (black line). This decay length strongly depends on the Gilbert damping and the inhomogeneous linewidth broadening of magnetic waveguides. The inset of Fig. 3(b) shows the calculated frequency shift of the FMR mode ($\triangle f_{FMR}$) as a function of the precession angle ($\theta$). A noticeable frequency shift is observed once the precession angle is larger than 10°. In general, Fig. 3 indicates the physical origin of the wavelength conversion due to the nonlinear shift. Two features are observed: 1) the FMR mode is directly excited under the antenna due to the large precession angle and the microwave energy is efficiently pumped into the magnonic domain; 2) the wavelength conversion from a large wavelength to a nanometre wavelength is completed at the edge of the antenna within a small conversion distance.

**High group velocity of exchange spin waves.** In the previous sections, the large nonlinear frequency shift was observed using μBLS spectroscopy. The semi-analytical theory and micromagnetic simulations show that the large nonlinear shift causes a wavelength conversion and leads to the emission of short-wavelength spin waves. Here, we show the experimental evidence of this conversion by measuring the large group velocities of the generated exchange spin waves, using time-resolved μBLS spectroscopy. A microwave pulse with a length of 600 ns and a repetition time of 1000 ns is applied to the antenna to excite a spin-wave pulse. As described above, a frequency shift of 2.1 GHz can be achieved by continuously sweeping the microwave signal with frequency steps of 10 MHz. In this case, the dispersion curve only needs to be shifted by a small frequency gap for each step, which does not require too much power. However, due to the large gap between the FMR frequency and the excitation frequency, it is difficult to directly excite high frequency (e.g., ≥6.4 GHz) spin waves with a single pulse. The straightforward way to obtain a high-frequency pulse would be to dramatically increase the input microwave power. However, to excite a high-frequency spin-wave pulse more elegantly and with relatively low microwave power and low heating effects, we introduce an additional low-frequency (e.g., ≤6.2 GHz), 50 ns-wide trigger pulse with the same repetition time at the beginning of the excitation frequency pulse. The trigger frequency has a small gap to the FMR frequency and can be excited with low power. As soon as the trigger signal is excited, the FMR frequency of the dispersion curve underneath the antenna is temporarily shifted to the trigger frequency and the gap between the FMR frequency and main excitation frequency is reduced. After that, the dispersion curve continuously shifts to the targeted excitation frequency. The trigger signal acts as a staircase, dividing a large frequency gap into two small gaps that can be easily overcome by applying a relatively small power.

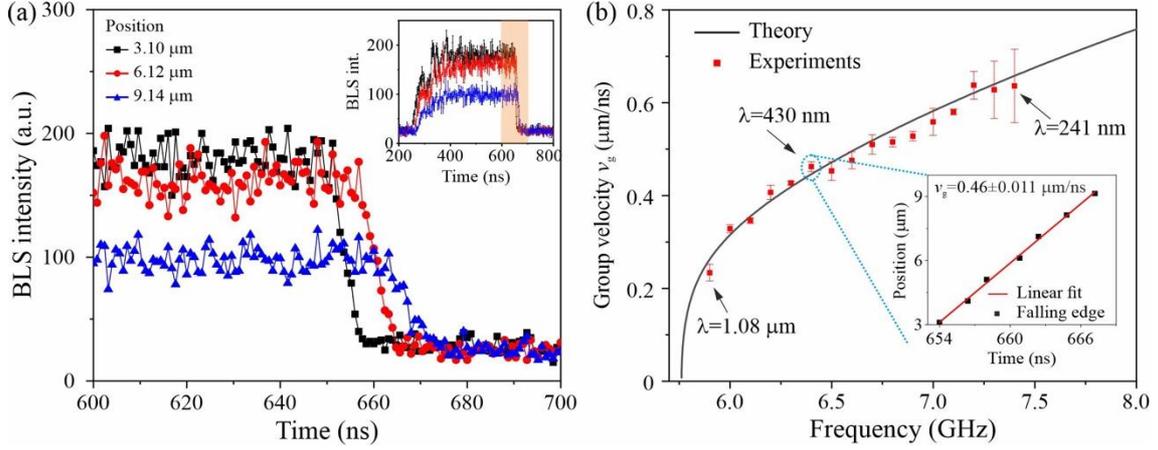

Figure 4. **High group velocity of exchange spin waves.** (a) The inset shows the full range of spin-wave packets at 6.4 GHz carrier frequency measured at different positions along the waveguide using time-resolved µBLS spectroscopy. The main panel shows a zoom-in region of the falling edge of the packets from 600 ns to 700 ns. (b) Spin-wave group velocity as a function of frequency. The solid line is the theoretical calculation. Several wavelengths are marked. The inset shows a linear fit of the measurement position as a function of the falling edge time of the spin-wave packets.

The inset of Fig. 4(a) shows the full range of spin-wave packets of frequency 6.4 GHz measured at different positions along the waveguide and excited by the trigger method with a trigger pulse (50 ns) with a frequency of 5.9 GHz (the time decay between the trigger and the main pulse is around 300 ns). The main panel of Fig. 4(a) shows a zoom-in region of the falling edge of the packets from 600 ns to 700 ns, indicated by the orange rectangular region in the inset. The BLS intensity is extracted by integrating the signal around 6.4 GHz, which excludes the contribution of the low-frequency trigger signal. A linear fit of the space-time dependency of the falling edge of the packet, as shown in the inset of Fig. 4(b), gives the averaged group velocity $v_g$ during the measurement distance. The experimental results (red dots) agree well with the theoretical prediction (black line), as illustrated in the main panel of Fig. 4(b). The increase in group velocity from 0.23±0.02 µm/ns ($f$=5.9 GHz, $\lambda$=1.08 µm) to 0.64±0.08 µm/ns ($f$=7.4 GHz, $\lambda$=241 nm) is observed, thus a direct evidence of the exchange character of the generated spin waves and proof of principle of the conversion mechanism described above. Note that we are able to detect spin-waves with wavenumbers above the theoretical limit of our µBLS, which is $k_{max}$ ~ 21 rad/µm ($\lambda_{min}$=300 nm for 457 nm laser wavelength and objective lens with N.A.=0.75) [39,40]. This can be explained by the fact that the 200 nm-wide YIG waveguide acts as a photonic nanoresonator that locally restricts the electromagnetic field and increases the range of accessible wavevectors [40] or the locally laser heating converts the exchange spin waves back to detectable dipolar spin waves.

**Deep exchange spin waves in CoFeB nano-waveguides.** In the following, we discuss the wavelength limitation of the proposed spin-wave excitation method. In the experiments, we successfully excite spin waves with frequencies up to 7.8 GHz (see Fig. 2(a)), which corresponds to a wavenumber up to $k$=30 rad/μm ($\lambda$=200 nm). However, this wavenumber is not the upper limit of the proposed method. As mentioned earlier, the wavelength strongly depends on the nonlinear frequency shift. Equation (1) clearly shows that this shift is proportional to the nonlinear coefficient $T_k$ and the precession angle $\theta$. $T_k$ is proportional to the saturation magnetisation of the material, and the maximum achievable precession angle $\theta$ is limited by various multi-magnon instabilities or/and higher-order width mode excitation, which are both defined by the material parameters and the geometry of the waveguides. Based on this knowledge, another common magnonic material, CoFeB, which has a high saturation magnetisation and a large exchange constant, is selected to investigate the potential capabilities of the proposed method. The following typical parameters of CoFeB are used in analytical calculations and micromagnetic simulations: saturation magnetisation $M_s$=1250 kA/m, exchange constant $A$=15 pJ/m, and Gilbert damping $\alpha$=2×10$^{-3}$. The geometric size of the CoFeB waveguide was chosen to be 50 nm wide and 5 nm thick to obtain a wide frequency range of the single-mode dispersion curve. An external field of 2.55 T is applied out-of-plane to saturate the CoFeB waveguide. The needed external field can be dramatically reduced by using material systems with perpendicular magnetic anisotropy [44]. The micro-magnetically simulated dispersion curve in the linear region is shown in Supplementary Material and agrees well with the analytical calculation shown in Fig. 5(a) (black line). Due to the nanoscale width and large exchange interaction, the minimum frequency of the third width mode was raised to 51.6 GHz (see Supplementary Materials). The FMR frequency of the first width mode in the linear region is 34.4 GHz, so a large single-mode frequency window of 17.2 GHz is observed. Similar to the experiments, a microwave signal is applied on a 2 μm-wide antenna and swept continuously from 34.4 GHz to 51 GHz with a frequency step of 0.2 GHz. Figure 5(b) shows the simulated width-averaged spin-wave amplitude ($m_x$) along the propagation waveguide for a frequency of 51 GHz. It clearly illustrates that the FMR mode underneath the antenna is excited with an averaged precession angle of around 51° and then is efficiently converted to exchange spin waves. A short wavelength of ~60 nm is observed directly near the antenna, and the minimum wavelength of 45 nm is reached at around 6 μm away from the antenna. The wavelength conversion rule is represented by the dashed blue line in Fig. 5(a): the wavenumber converts from the FMR mode to deep exchange spin waves with the wavenumber up to $k_3$=138 rad/μm ($\lambda_3$=45 nm). In principle, the wavelength can be further reduced with the proposed method by optimising the material parameters and the geometry of the waveguides. Furthermore, the bistability and the self-normalising property of the spin wave source are also available for the CoFeB systems, as the underlying principle is the same as for the previously discussed YIG systems.

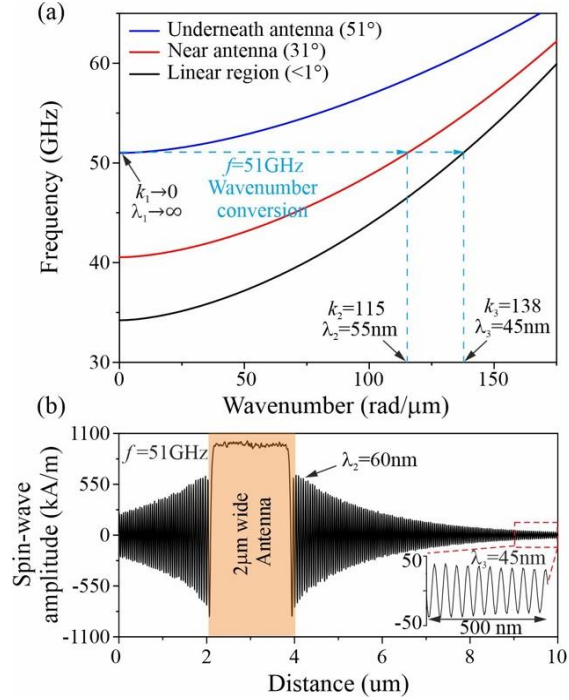

Figure 5. **Deep exchange spin waves in a nanoscale CoFeB waveguide**. (a) The calculated dispersion curves of the 50 nm wide and 5 nm thick CoFeB waveguide for different precession angles. The dashed blue line shows the wavenumber conversion for a spin-wave frequency of 51 GHz. (b) The simulated width-averaged spin-wave amplitude ($m_x$) as a function of distance in the CoFeB waveguide.

**Discussion**

In a 200 nm-wide YIG waveguide, we have experimentally observed a nonlinear positive frequency shift of FVSWs up to 2.1 GHz, which corresponds to a huge precession angle of around 55°. This large positive frequency shift is achieved due to the out-of-plane magnetisation geometry and a large region with a single-mode spin-wave dispersion curve in nanoscale waveguides. Based on these findings, we have proposed a universal method to excite a whole range of spin waves, from microscale dipolar spin waves to nanoscale exchange spin waves, by using this deep positive nonlinear frequency shift. In the experiments, spin waves with wavelengths ranging from a few micrometres to 200 nm are successfully excited using the proposed method. In addition, the power of the excited spin-wave is independent of the microwave input power, which opens access to significant simplification of the designs of the magnonic circuits. Further simulations reveal that in nanoscale CoFeB waveguides deep exchange spin waves with a wavelength down to 45 nm can be excited using the same mechanism. The proposed method removes the wavelength limitations imposed by the size of inductive antennas, increases the excitation efficiency of exchange waves and enables direct on-chip integration. Although we consider only FVSW geometry here, the method is predicted to work in other cases provided that the nonlinear frequency shift is positive (in-plane magnetised waveguides with large perpendicular anisotropy, for instance). We sincerely believe that this method will give new impetus to the development of nanoscale integrated magnonic circuits.

**Methods**

**Nanoscale waveguide fabrication.** The YIG thin film is grown on top of a 500 μm-thick (111) gadolinium gallium garnet (GGG) substrate by liquid phase epitaxy (LPE) [45]. The parameters of the unstructured thin film have been characterised by stripline vector-network-analyser ferromagnetic resonance spectroscopy and BLS spectroscopy and yield a saturation magnetisation of $M_s = (140.7 \pm 2.8)$ kA/m, Gilbert damping parameter $\alpha = (1.75 \pm 0.08) \times 10^{-4}$, inhomogeneous linewidth broadening $\mu_0 \Delta H_0 = (0.18 \pm 0.01)$ mT, and exchange constant $A_{ex} = (4.22 \pm 0.21)$ pJ/m. These parameters are typical for high-quality thin YIG films [35,45]. Nanoscale YIG waveguides were fabricated using a Cr/Ti hard-mask and ion beam milling procedure, described in detail in Ref. [35].

**BLS measurements.** A single-frequency laser with a wavelength of 457 nm is used, focused on the sample using a microscope objective (magnification 100× and numerical aperture N.A.=0.75). The laser power of 2.6 mW is focused on the sample. The BLS detection efficiency decreases with the increase of the spin-wave wavenumber. In order to increase the BLS signal, the laser power is increased to 6.6 mW for high frequency ($f$>6.8 GHz) group velocity measurements in Fig. 4(b). A uniform out-of-plane external field of 330 mT is provided by a NdFeB permanent magnet with a diameter of 70 mm. Microwave signals with different powers were applied to the antenna to excite spin waves.

**Micromagnetic simulations.** The micromagnetic simulations were performed by the GPU-accelerated simulation package Mumax$^3$, including both exchange and dipolar interactions, to calculate the space- and time-dependent magnetisation dynamics in the investigated structures [46]. The parameters of a nanometre-thick YIG film were used [35]: saturation magnetisation $M_s = 1.407 \times 10^5$ A/m, exchange constant $A = 4.2$ pJ/m. The Gilbert damping is increased to $\alpha = 5 \times 10^{-4}$ to account for the inhomogeneous linewidth which cannot be directly plugged into Mumax$^3$ simulations. The Gilbert damping at the end of the device was set to exponentially increase to 0.5 to avoid spin-wave reflection. The mesh was set to $10 \times 10 \times 44$ nm$^3$ (single layer along the thickness) for YIG waveguide. An external field $B_{ext} = 330$ mT is applied along the out-of-plane axis ($z$-axis as shown in Fig. 1) and thus sufficient to saturate the structure in this direction.

The typical parameters of CoFeB were used: saturation magnetisation $M_s=12.5\times10^5$ A/m, exchange constant $A=15$ pJ/m, and the Gilbert damping $\alpha=2\times10^{-3}$. The mesh was set to $5\times5\times5$ nm$^3$. An external field $B_{ext}=2.55$ T is applied to out-of-plane.

To excite propagating spin waves, we first calculate the Oersted field distribution of a 2 μm wide strip antenna with current of 25 mA in the magneto-static approximation and plug it into Mumax$^3$ with a varying microwave frequency $f$. The $M_y(x,y,t)$ of each cell was collected over a period of 100 ns and recorded in 25 ps intervals. The fluctuations $m_y(x,y,t)$ were calculated for all cells via $m_y(x,y,t) = M_y(x,y,t) – M_y(x,y,0)$, where $M_y(x,y,0)$ corresponds to the ground state. The spin-wave dispersion curves were calculated by performing a two-dimensional fast Fourier transformation of the fluctuations.

**Calculation of linear and nonlinear dispersion curves.**

**Linear forward volume spin waves dispersion curve in nanoscale waveguide.** In nanoscale waveguides, the width profile of spin waves often becomes nonuniform with effective wavenumber along the width strongly dependent on the waveguide geometry and material [34]. In the FVSW case, this nonuniformity is defined by nonuniform static demagnetisation fields, leading to a partial pinning of spin waves at the lateral edges of the waveguide. To the date, there is no analytical theory describing this general case. Therefore, we use a numerical approach for the spin wave profile and dispersion calculation. The spin-wave profile $\boldsymbol{m}_y(y)$, $k=k_x$, and frequency can be found by the following equation [34]:

$$-i\omega_k \boldsymbol{m}_k = \boldsymbol{\mu} \times \widehat{\boldsymbol{\Omega}}_k * \boldsymbol{m}_k$$

where $\boldsymbol{\mu} = \boldsymbol{e}_z$ is the static magnetisation direction, $\widehat{\Omega}_k$ is the tensorial operator

$$\widehat{\boldsymbol{\Omega}}_k = \gamma B(y)\hat{I} + \omega_M \widehat{N}_k$$

with $B(y)$ being the profile of the static internal field and $\widehat{N}_k$ – the operator of magnetic self-interactions (exchange and dipolar interaction):

$$\widehat{N}_k * \boldsymbol{m}_k(y) = \lambda^2 \left[k^2 - \frac{d^2}{dy^2}\right]\boldsymbol{m}_k(y) + \int \widehat{\boldsymbol{G}}_k(y-y') \cdot \boldsymbol{m}_k(y')dy'$$

where $\widehat{\boldsymbol{G}}_k$ is the tensorial magnetostatic Green's function [47].

The principal difference from the case of backward volume spin waves, studied in [34], is that the static internal field is essentially nonuniform and is calculated as

$$B(y) = B_e - \mu_0 M_s \int G_0^{(zz)}(y-y')dy'$$

where we assume uniform static magnetisation and external field applied in the out-of-plane direction, $\boldsymbol{B}_e = B_e \boldsymbol{e}_z$.

**Nonlinear frequency shift.** Since the mode profile is nonuniform, it is hard to apply standard Hamiltonian formalism for the nonlinear spin-wave dynamics and derive an analytical expression for the nonlinear frequency shift coefficient. Instead, we used the recently developed vectorial Hamiltonian formalism [48,49]. The coefficient of the nonlinear frequency shift is derived as

$$T_k = \frac{\omega_M}{4w} \int dy [((\boldsymbol{m}_k^* \cdot \boldsymbol{m}_k^*)\boldsymbol{\mu} \cdot \widehat{\boldsymbol{N}}_{2k} * \boldsymbol{\mu}(\boldsymbol{m}_k \cdot \boldsymbol{m}_k) + c.c.) + 4|\boldsymbol{m}_k|^2 \boldsymbol{\mu} \cdot \widehat{\boldsymbol{N}}_0 * \boldsymbol{\mu}|\boldsymbol{m}_k|^2$$
$$- (2|\boldsymbol{m}_k|^2 \boldsymbol{m}_k^* \cdot \widehat{\boldsymbol{N}}_k * \boldsymbol{m}_k + (\boldsymbol{m}_k^* \cdot \boldsymbol{m}_k^*)\boldsymbol{m}_k \cdot \widehat{\boldsymbol{N}}_k * \boldsymbol{m}_k + c.c.)]$$

Here $w$ is the width of the waveguide, the mode frequency changes with its amplitude as $\omega_k(c_k) = \omega_{k,0} + T_k|c_k|^2$, and the mode profiles are normalised to 1, i.e., $(1/w) \int i\, \boldsymbol{m}_k^* \cdot \boldsymbol{\mu} \times \boldsymbol{m}_k dy = 1$. The relation between mode amplitude and real magnetisation is given by

$$\boldsymbol{M}/M_s = \left(1 - \frac{|s|^2}{2}\right)\boldsymbol{\mu} + \sqrt{1 - \frac{|s|^2}{4}}\, \boldsymbol{s},$$

where

$$\boldsymbol{s}(\boldsymbol{r}, t) = c_k(t)\boldsymbol{m}_k(y)e^{-ikx} + c.c.$$

In our case the magnetisation precession is close to circular, and it is possible to approximately relate the width-averaged precession angle $\bar{\theta}$ to the mode amplitude as

$$\sin\bar{\theta} = \sqrt{2 - |c_k|^2}|c_k|, \text{ i.e., } |c_k|^2 = 1 - \cos\bar{\theta}.$$

**Comparison of energy transmission.** From the simulations, we can calculate the power carried by a propagating spin wave and power dissipated under the antenna region. The energy density of spin waves can be calculated as:

$$W_{sw} = \frac{M_s}{\gamma}\left(\omega_k|c_k|^2 + \frac{1}{2}T_k|c_k|^4\right)$$

where $\omega_k$ is the linear spin-wave frequency at given wavenumber $k$, $T_k$ is the nonlinear frequency shift, and spin-wave amplitude is $|c_k|^2 = 1 - \cos\theta$.

The power carried by the propagating spin wave is $P_{sw} = whv_g W_{sw}$, where $w$ and $h$ are the width and thickness of waveguide, $v_g$ is the velocity of spin waves, which can be estimated as the group velocity calculated from nonlinear dispersion curve at given $k$ (although in the case of strongly nonlinear spin waves these quantities could be different). The power dissipated under the antenna is $P_G = 2\Gamma wh \int W_{sw}(x)dx$, where the integration goes over the antenna width and the $\Gamma$ is the damping rate. In our case, $P_G$ has sense of parasitic losses, while $2P_{sw}$ is useful power (multiplier 2 corresponds to 2 counter-propagating wave from antenna). Then, we can estimate energy transmission ratio $T_R$ as

$$T_R = \frac{2P_{sw}}{2P_{sw} + P_G}$$

**Data availability**

The data that support the plots presented in this paper are available from the corresponding authors upon reasonable request.

**Code availability**

The code used to analyse the data and the related simulation files are available from the corresponding author upon reasonable request.

**Acknowledgements**

The project is funded by the Austrian Science Fund (FWF) via Grant No. I 4696-N (Nano-YIG) and Grant No. F65 (SFB PDE), the Vienna Science and Technology Fund (WWTF) project MA16-066 "SEQUEX", the European Research Council (ERC) Starting Grant 678309 MagnonCircuits and the Deutsche Forschungsgemeinschaft (DFG, German Research Foundation) – 271741898 and TRR 173 - 268565370 ("Spin + X", Project B01). R.V. acknowledges support by National Academy of Sciences of Ukraine, project No. 0122U020220. O.W. was supported by Brno PhD talent scholarship and acknowledge support from the project Quality Internal Grants of BUT (KInG BUT), Reg. No. CZ.02.2.69/0.0/0.0/19\_073/0016948, which is financed from the OP RDE.

**Author Contributions**

Q. W. conceived the idea and carried out micromagnetic simulations and performed BLS measurements with support from M. S., B. H. and K. L.. R. V. developed the analytical theory and performed the theoretical calculations. B. H. and K. D. fabricated the nanoscale YIG waveguides. C. D. grew the YIG film. O. W. and M. U. discussed the results of BLS measurements for forward volume spin waves. A. V. C. and P. P. led this project. Q. W. wrote the manuscript with the help of all the coauthors. All authors contributed to the scientific discussion and commented on the manuscript.

**Competing Interests**

The authors declare no competing interests.

**Correspondence** and requests for materials should be addressed to Q. W.

# Supplementary information
# Deeply nonlinear excitation of self-normalised exchange spin waves

Qi Wang[1,2,3], Roman Verba[4], Björn Heinz[5], Michael Schneider[5], Ondřej Wojewoda[6], Kristyna Davidkova[6], Khrystyna Levchenko[1], Carsten Dubs[7], Norbert J. Mauser[2,3], Michal Urbánek[6], Philipp Pirro[5], Andrii V. Chumak[1,2]

[1] Faculty of Physics, University of Vienna, Vienna, Austria

[2] Research Platform Mathematics-Magnetism-Materials, c/o Faculty of Math, University of Vienna, Vienna, Austria

[3] Wolfgang Pauli Institute, Vienna, Austria

[4] Institute of Magnetism, Kyiv, Ukraine

[5] Fachbereich Physik and Landesforschungszentrum OPTIMAS, Technische Universität Kaiserslautern, Kaiserslautern, Germany

[6] CEITEC BUT, Brno University of Technology, Brno, Czech Republic

[7] INNOVENT e.V., Technologieentwicklung, Jena, Germany

In the supplementary information, Section 1 discusses the comparisons between the simulated dispersion curves, the classical spin-wave dispersion curve theory and the semianalytical theory. In Section 2, the simulated spin-wave intensity as a function of excitation frequency is presented. In Section 3, the influence of the Joule heating on the nonlinear frequency shift is discussed. The contribution of nonlinearity to the energy conversion from FMR mode to exchange spin waves is discussed in Section 4.

1. **Forward volume spin waves dispersion curve in nanoscale waveguides**

Dispersion characteristic is fundamental information for magnonic studies. The dipole-exchange spin-wave spectrum theory for ferromagnetic films, developed by Kalinikos and Slavin, is a classical theory which has been widely used in the calculation of the dispersion curve in macro/micro-scale waveguides [S1]. However, this theory was developed with a uniform static demagnetisation field approximation, which is not applicable to the nanoscale waveguides with nonuniform internal field in our studies. Here, we present a semianalytical theory that accounts for the nonuniform static demagnetisation field. This nonuniformity leads to partial pinning of spin waves at the lateral edges of waveguide and changes the propagation characteristics of spin waves, i.e., the dispersion curve. Figure 1 shows the comparison between the simulated dispersion curve (colormap), the classical theory (red line) and the presented semianalytical theory (black dashed line). The parameters of simulation and calculation are the same as those used in the main text. It can be clearly seen that the semianalytical theory (black dashed line) was able to well reproduce the simulated results. However, there is a discrepancy between the classical theory (red line) and simulated dispersion curve (colormap). Therefore, the nonuniformity has to be considered during the calculation of dispersion curve for the nanoscale waveguides. Furthermore, the simulated dispersion curves also show that the higher width modes have been pushed up significantly in frequency due to waveguide's nanoscale width and large exchange contribution.

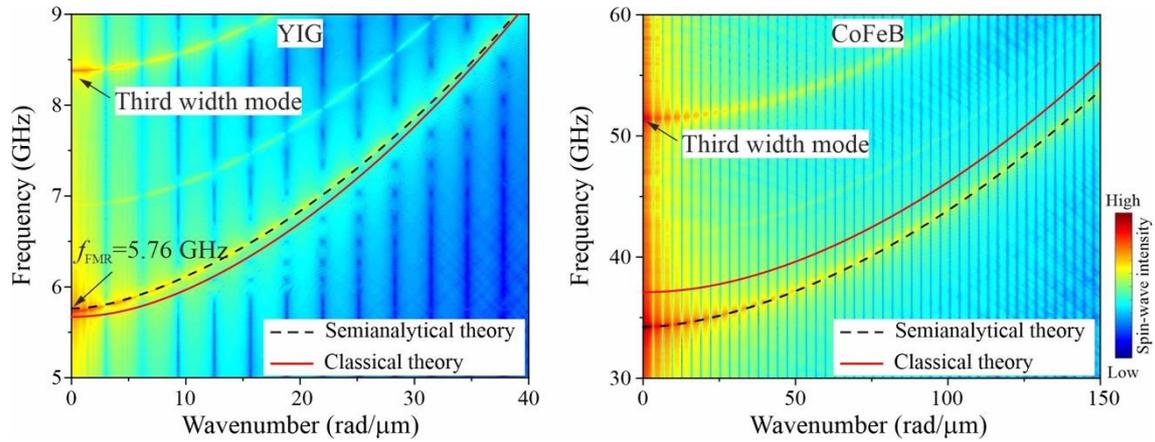

Fig. S1 Simulated dispersion curves (colormap) of (a) a 200 nm wide 44 nm thick YIG waveguide with 330 mT out-of-plane magnetic field and (b) a 50 nm wide 5 nm thick CoFeB waveguide with 2.55 T out-of-plane magnetic field. The red solid line and black dashed line represent the classical theory prediction and the semianalytical theory.

## 2. Simulated spin-wave intensity as a function of excitation frequency

In the main text, we attribute the decrease in the BLS intensity in the high-frequency range (above 6.4 GHz) to the reduced detection efficiency of μBLS for short wavelengths. To verify this, we performed micromagnetic simulations similar to the experiments, where the excitation frequency sweeps from 5.6 GHz to 7 GHz (or vice versa) with a step size of 20 MHz. Figure S2 shows the simulated spin-wave spectra with similar foldover and bistability behaviour. More importantly, the simulated spin-wave intensity increases monotonically with the increase of frequency, which confirming our statement in the main text that the decrease in the BLS intensity in the high-frequency range is due to the decrease of the detection efficiency of μBLS for short wavelengths.

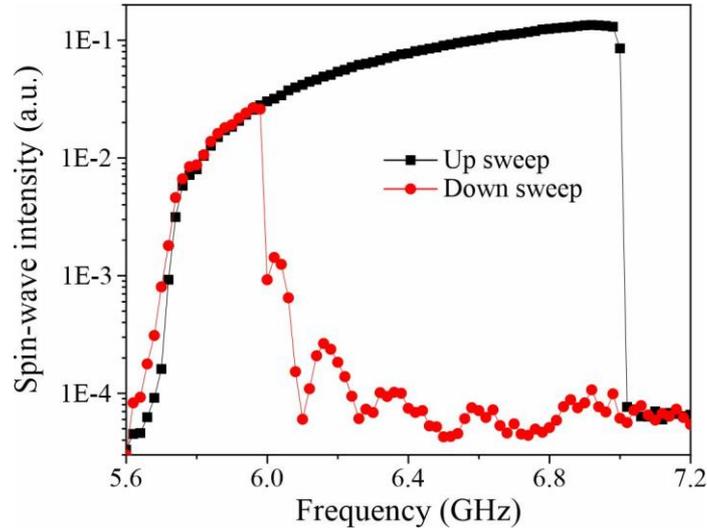

Fig. S2 Simulated spin-wave intensity as a function of excitation for up- and down-frequency sweep.

## 3. The influence of Joule heating on nonlinear frequency shift

In Fig. 3(a) of the main text, a large frequency shift is observed by using continuous wave (CW) excitations. To investigate the effect of the Joule heating produced by the microwave current on the frequency shift, pulse excitations with different duration times are used. The power and repetition time of the pulse are fixed at 12 dBm and 1000 ns, respectively. The pulse duration varies from 900 ns to 300 ns as shown in Fig. S3. The black dot line shows the reference signal obtained by down-sweep CW excitation (from high frequency to low frequency). In all cases, a clear frequency jump point is observed, which is a good "sensor" for estimating the influence of the Joule heating. The frequency of the jump point slightly decreases by 20 MHz from 5.79 GHz (CW) to 5.75 GHz (300 ns pulse). The contribution of Joule heating to the frequency shift can be neglected compared to the total frequency shift of 2.1 GHz. Therefore, the frequency shift obtained in the main text is mainly due to the nonlinear phenomena.

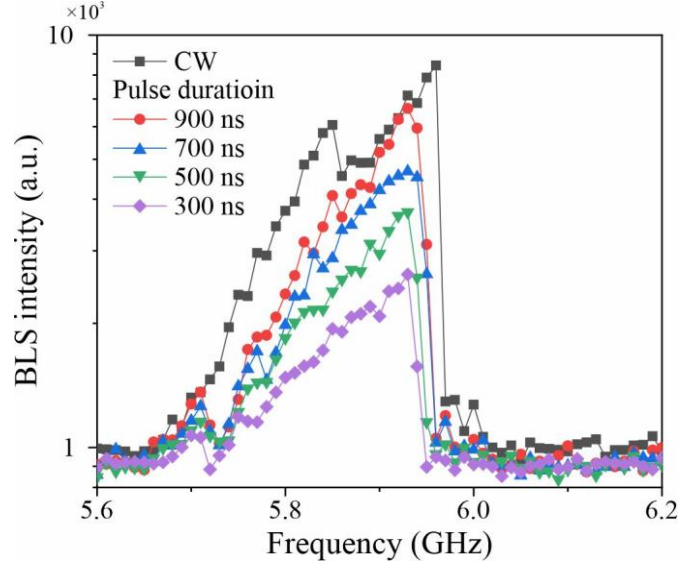

Fig. S3 The µBLS intensity as a function of excitation frequency for continue wave (black dot line) and pulse excitation with different pulse durations.

## 4. The contribution of nonlinearity for the energy conversion

In the main text, the energy conversion from FMR mode to exchange spin waves is estimated to be more than 80% by calculating the power of propagating spin waves and the power dissipated under the antenna region. To verify the contribution of the nonlinearity to the energy conversion process, a spatial modulated external field is constructed and inserted into Mumax3 to simulate the wavenumber conversion in the linear case.

Firstly, we take the internal field from Fig. 3(a) and increase it by ~130 mT to compensate for the demagnetisation field, and plug it into Mumax$^3$ as a spatial modulated external field. In this case, the FMR frequency under the antenna is around 7.55 GHz in the linear region. The top panel of Fig. S4 shows the spatial dependent internal field extracted from Mumax$^3$. Secondly, a very small Oersted field generated by the 2 µm antenna is used to excite the linear spin waves with precession angles below 1°. The middle panel of Fig. S4 shows the snapshot of the simulated spin-wave amplitude $m_x$ as a function of propagation distance in the waveguide. The bottom panel shows the precession angle as a function of the propagation distance, where most of the energies are concentrated under the antenna. A similar wavelength conversion from FMR mode to exchange waves is observed, but with lower energy conversion efficiency. Table S1 summarized of the energy transmission estimated using the way mentioned in the Methods. It shows that the energy transmission ratio $T_R$ is 83% for nonlinear case and 55% for linear case. If we compare the parasitic losses, i.e., $L_p = 1 - T_R$, the nonlinear mechanism is around 2.6 times more efficient than linear one. Finally, in the main text, we concluded that despite the similarity, such conversion is much less efficient when the spin-wave dispersion shift is determined not by nonlinearity, but by spatial modulation of the external field or magnetic parameters, since a spin-wave is partially reflected from such a sharp boundary.

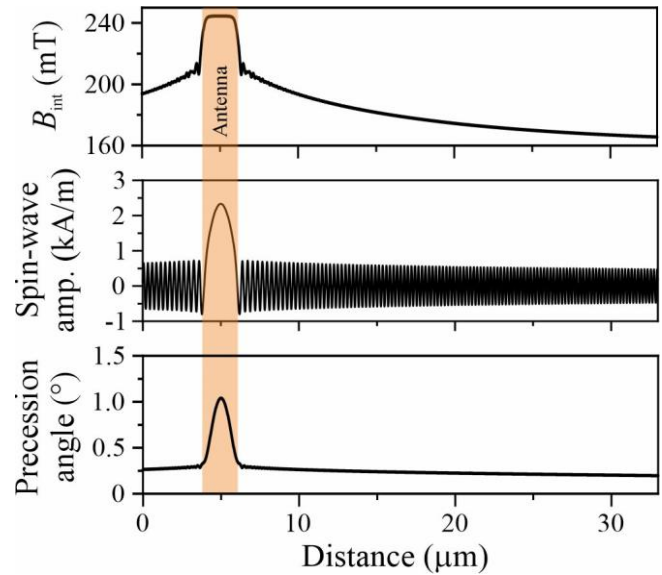

Fig. S4 The internal field (top panel), the spin-wave amplitude (middle panel), and the precession angle (bottom panel) as a function of distance for linear excitation.

Table S1. The summary of the energy transmission

|  | Averaged SW energy density under antenna $<W_{sw}>$ (J/m$^3$) | SW energy density close to antenna (at 500 nm distance) $<W_{sw}>$ (J/m$^3$) | $v_g$ (m/s) | PG | $2P_{sw}$ | $T_R$ % |
|---|---|---|---|---|---|---|
| Nonlinear case | 15591 | 10270 | 348 | 13 nw | 63 nW | 83 |
| Linear case | 3.61 | 0.53 | 408 | 3.1 pW | 3.8 pW | 55 |

**Supplementary References:**